\documentclass[letterpaper, 10 pt, conference]{ieeeconf}  

\IEEEoverridecommandlockouts                              

\usepackage{graphics} 
\usepackage{epsfig} 
\usepackage{mathptmx} 
\usepackage{times} 
\usepackage{amsmath} 
\usepackage{amssymb}  
\usepackage{color, algorithm, algpseudocode, multirow, array, float}
\usepackage{booktabs, multicol, multirow, bbding}

\title{\LARGE \bf
Snake with Shifted Window: Learning to Adapt Vessel Pattern for OCTA Segmentation
}

\author{Xinrun Chen$^{1}$, Mei Shen$^{2}$, Haojian Ning$^{1}$, Mengzhan Zhang$^{2}$, Chengliang Wang$^{1*}$ and Shiying Li$^{2*}$    
\thanks{----------------------------------------------------------------------------------------}
\thanks{* Correspondence:}%
\thanks{\ \ \ \ \ \ \ Chengliang Wang | Wangcl@cqu.edu.cn}%
\thanks{\ \ \ \ \ \ \ Shiying Li | shiying\_li@126.com}%
\thanks{${1}$. Chongqing University, College of Computer Science, Chongqing, China}%
\thanks{${2}$. Xiang’an Hospital of Xiamen University, Department of Ophthalmology, Xiamen, China}%
}

\begin{document}

\maketitle
\thispagestyle{empty}
\pagestyle{empty}

\textit{\small{Statement: This work has been submitted to the IEEE for possible publication. Copyright may be transferred without notice, after which this version may no longer be accessible.}}
\\

\begin{abstract}

Segmenting specific targets or structures in optical coherence tomography angiography (OCTA) images is fundamental for conducting further pathological studies. The retinal vascular layers are rich and intricate, and such vascular with complex shapes can be captured by the widely-studied OCTA images. In this paper, we thus study how to use OCTA images with projection vascular layers to segment retinal structures. To this end, we propose the SSW-OCTA model, which integrates the advantages of deformable convolutions suited for tubular structures and the swin-transformer for global feature extraction, adapting to the characteristics of OCTA modality images. Our model underwent testing and comparison on the OCTA-500 dataset, achieving state-of-the-art performance. The code is available at: https://github.com/ShellRedia/Snake-SWin-OCTA.

\end{abstract}

\section{Introduction}
\label{Sec_Intro}

\begin{figure}[h]
  \centering
  \includegraphics[width=1\linewidth]{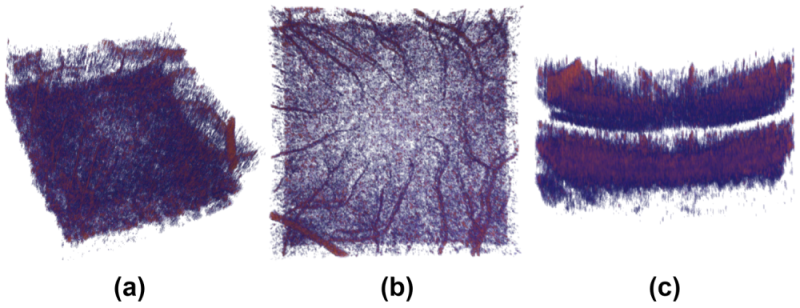}
  \caption{OCTA scanned 3D volume in various views: (a) perspective; (b) en-face; (c) B-scan.}
  \label{Fig_OCTA_Scan}
\end{figure}

Retinal vascular analysis forms the foundation for various ophthalmic diseases. With the advancement of fluorescein angiography and other imaging techniques in vascular imaging, the visualization of ocular vasculature has progressively improved \cite{kalra2022optical}. Optical Coherence Tomography Angiography (OCTA),  as a non-invasive and rapid imaging technique, allows the generation of cross-sectional, in vivo images of the dynamic microvasculature of the choroid and retina. OCTA scan produces en-face images and cross-sectional B-scans with blood flow signals, offering a comprehensive view of retinal and choroidal vasculature. En-face OCTA provides a top-down view of vascular layers, while B-scans offer structural and flow information for precise depth localization of vascular abnormalities \cite{javed2023optical}. In essence, the OCTA scan generates three-dimensional (3D) data, as illustrated in Figure \ref{Fig_OCTA_Scan}.

Ocular diseases, marked by conditions like ischemia and inflammation, can prompt neovascularization in response to a damaged retina or choroid, compensating for reduced oxygen levels \cite{tun2023complete}. This determines the significance of analyzing vessel morphology and distribution in OCTA images. In OCTA-related research, there is considerable focus on the segmentation of retinal vessels (RV) and the foveal avascular zone (FAZ)\cite{preity2023automated}. FAZ is a specialized avascular region at the macular margin that is crucial for central vision and can expand due to capillary dropout. This enlargement may indicate macular ischemia, potentially causing vision impairment and contributing to conditions like diabetic retinopathy (DR), age-related macular degeneration, and glaucoma \cite{pang2023foveal}. Some ocular diseases, such as DR, have a progression that impacts retinal arteries and veins to varying extents. Therefore, artery-vein (AV) segmentation also provides valuable information for the detection and diagnosis of diseases \cite{abtahi2022mf}.

Based on the characteristics of the ocular OCTA sample, we propose a model named SSW-OCTA (dynamic Snake convolution with Shifted Window transformer mechanism for OCTA segmentation) for the en-face image segmentation. The main contributions are summarized as follows:

(1) The proposed model synergizes dynamic snake convolution and attention mechanism for improved vessel feature extraction in OCTA enface 2D images, which achieves state-of-the-art (SOTA) RV and capillary segmentation even with a lightweight version.

(2) Two types of model architectures are designed to propose an appropriate feature fusion method, and key hyperparameters are tested in ablation experiments to offer references for different segmentation targets.
\section{Related Work}
\label{Sec_Related}

\subsection{OCTA segmentation models}

OCTA segmentation models are categorized as either 2D or 3D based on input dimensions. The output is 2D images of a region or structure in either an en-face or B-scan view. Due to the complexity of vascular distribution, the 3D annotations are hard to create, which makes related prediction models rare. 

Given the larger size of 3D inputs, transitional methods are employed to convert them into 2D feature maps, allowing for concurrent feature extraction. To achieve this, IPN incorporates a projection learning module for utilizing unidirectional pooling and feature selection \cite{li2020image}. PAENet utilizes a more adaptive pooling and quadruple attention module to exploit voxel dependencies along the projection direction, and RPS-Net adopted a parallel projection module design ~\cite{wu2021paenet,li2022rps}. 3D methods excel in utilizing complete sample information, benefiting accurate segmentation. Meanwhile, they have downsides like increased memory requirements and time-consuming data processing, which is not conducive to deployment or conversion lightweight.

2D methods use several representative projection images as model input. These images are generated by sampling the retina along the projection direction after anatomical layer segmentation. With the maturity of general 2D image processing methods, related research exhibits greater diversity in methodologies. The TCU-Net, OCT2Former, and StruNet improved the vision transformer (ViT) to extract the global features better because of the widespread distribution of vessels in OCTA images \cite{shi2023tcu, tan2023oct2former, ma2023strunet}. This ensures continuous retinal vessel segmentation, addressing issues like discontinuities or missing segments in the vasculature. Other methods, emphasizing efficiency, multi-scale, joint learning, and contrastive learning, have implemented various techniques and strategies, achieving favorable segmentation results on OCTA datasets ~\cite{wang2023db, zhu2022ovs, liang2021foveal,  ma2022retinal}. However, existing methods still have the potential for further improving the feature extraction efficiency for OCTA vasculature.

\subsection{Applicable techniques}

Models based on ViT are often designed to capture global features in OCTA images. However, the transformer structure lacks inductive bias, so substantial training data is necessary. Still, the public OCTA datasets with detailed annotations currently available remain limited in sample size. Swin-transformer adopts a hierarchical partitioned attention mechanism, introducing windowed self-attention for local computations \cite{liu2021swin}. This preserves ViT's ability to capture long-range dependencies and is more suitable for small sample sizes datasets. 

Dynamic snake convolution network (DSCNet) references deformable convolutions, adapting by altering convolutional kernel shapes to accurately capture the features of tubular structures by focusing on elongated and curved local structures \cite{qi2023dynamic}. A loss function named centerline Dice (clDice) loss is calculated based on the intersection of the segmentation mask of tubular structures and the morphological skeleton, ensuring topology preservation up to homotopy equivalence \cite{shit2021cldice}. These two techniques aid in the precise segmentation of vascular structures in OCTA images.
\section{Method} \label{Sec_Method}

\begin{figure}
  \centering
  \includegraphics[width=1\linewidth]{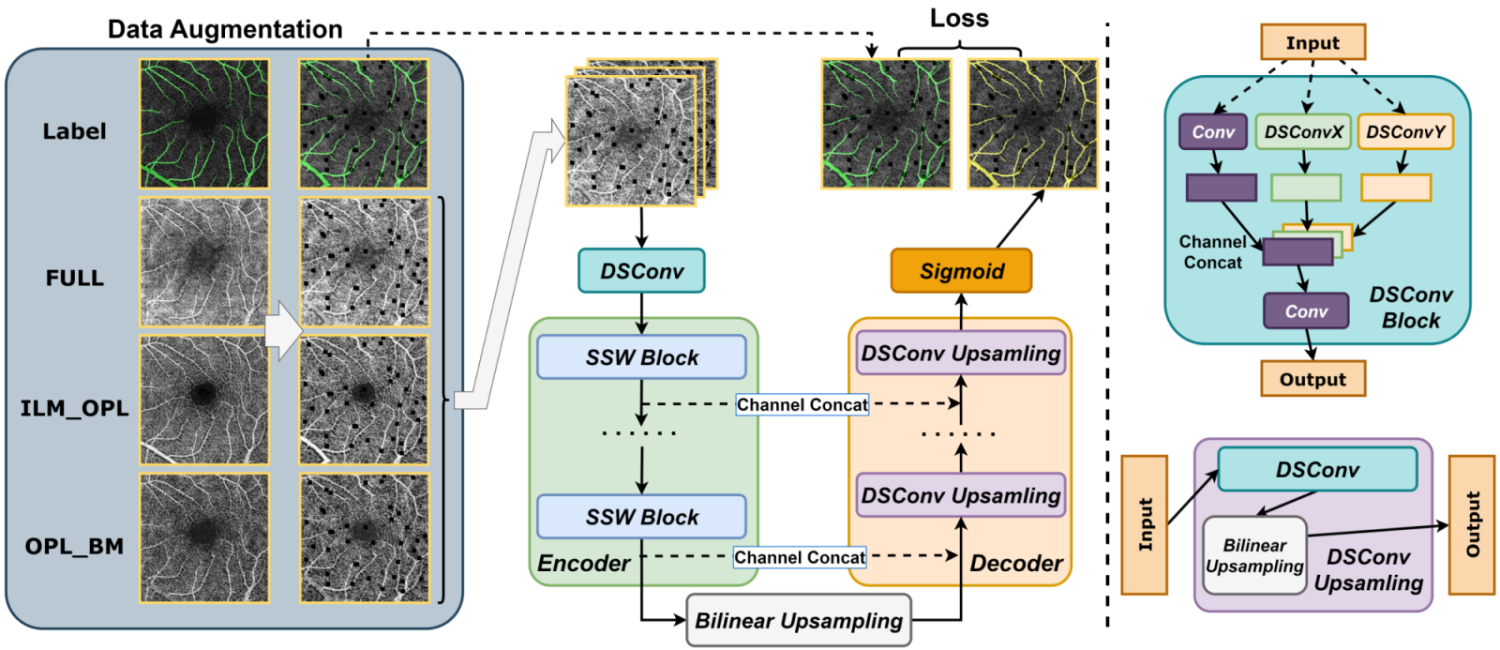}
  \caption{Schematic diagram of the SSW-OCTA model.}
  \label{Fig_SSW_OCTA}
\end{figure}

The overall structure of the SSW-OCTA model follows the paradigm of U-Net as shown in Figure \ref{Fig_SSW_OCTA}. In the encoder, swin-transformer and DSConv blocks represent two types of feature extraction modules. 


\subsection{Dynamic snake convolution block}

The purpose of the DSConv block is to deform the convolution kernel along the x-axis or y-axis, allowing it to capture specific spatial features in the image better. The deformation processing of the convolution kernel along the single-axis adopts an iterative adjustments strategy, ensuring that the receptive field focuses on the target. The DSConv kernel of the x-axis can be represented as:
\begin{align} \label{Eq_DSConv}
K_{i \pm c} = \begin{cases}
     (x_{i+c}, y_{i+c}) = (x_{i+c}, y_i+\sum_{i}^{i+c} \Delta y) \\
     (x_{i-c}, y_{i-c}) = (x_{i-c}, y_i+\sum_{i-c}^{i} \Delta y) \\
    \end{cases}
\end{align} 
where $\Delta = {\delta|\delta \in [ -1, 1 ]}$. 

\begin{figure}
  \centering
  \includegraphics[width=0.95\linewidth]{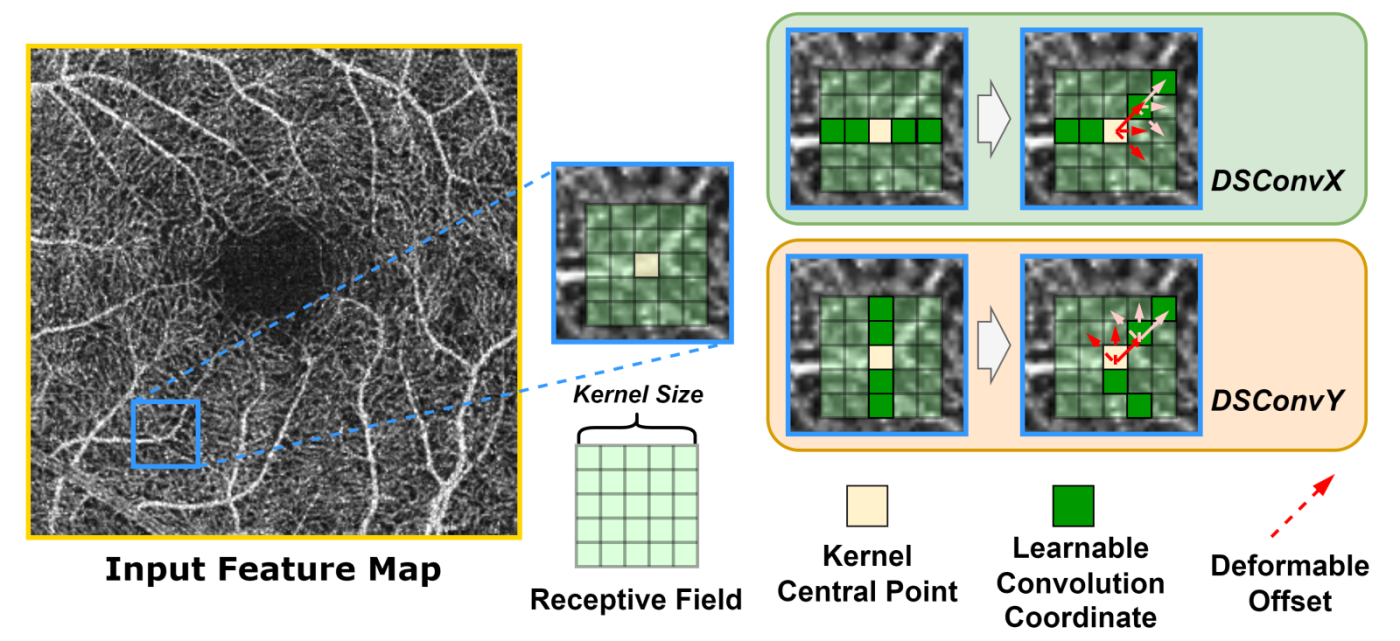}
  \caption{Deformation mechanism of dynamic snake convolution layer.}
  \label{Fig_DSConv}
\end{figure}

Similar to deformable convolution, $\Delta$ is a learnable parameter and should be a decimal, and $c$ is a hyperparameter representing the kernel size. The position of the convolution kernel requires integers, so the bilinear interpolation is adopted for conversion. The Formula \ref{Eq_DSConv} corresponds to the DSConv block in Figure \ref{Fig_SSW_OCTA}. The more detailed process is shown in Figure \ref{Fig_DSConv}. In the model implementation, DSConv blocks of the X and Y axes extract features with a traditional convolutional layer, which fuses features in three aspects.

\subsection{Blocks in swin-transformer}

\begin{figure}
  \centering
  \includegraphics[width=1\linewidth]{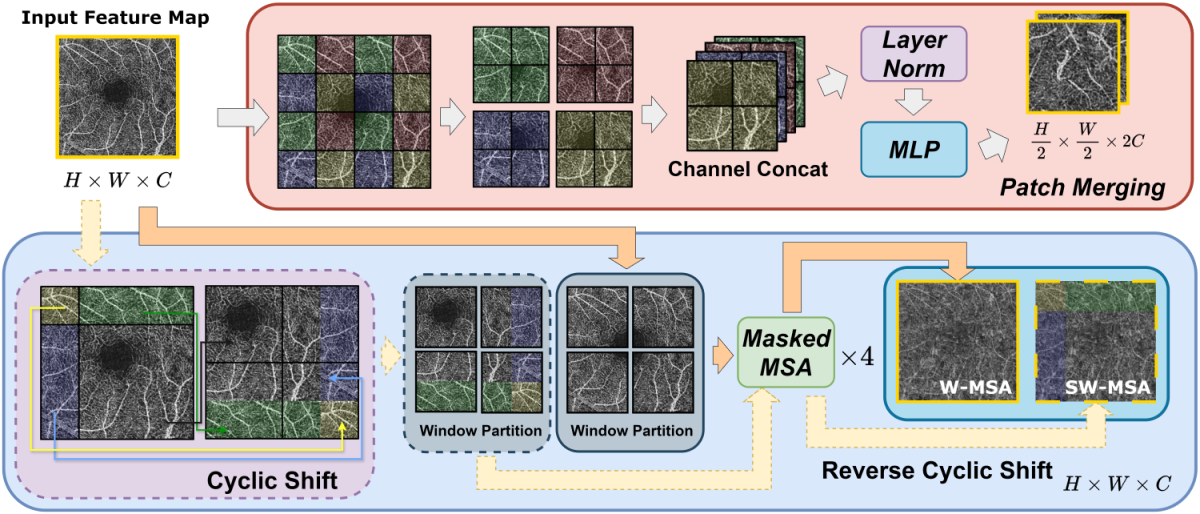}
  \caption{Patch merging and (S)W-MSA. The process of SW-MSA and W-MSA is similar. Both modules involve dividing the feature map into patches, placing them into different windows, computing masked multi-head self-attention within each window, and then reforming from patches. Compared to W-MSA, SW-MSA imports the ``cycle shift" step, enhancing the feature interaction between far-away patches.}
  \label{Fig_SWT_Blocks}
\end{figure}

The core modules in the swin-transformer are patch merging and the swin-transformer block. Patch merging is used in the ViT models to combine feature map patches, which is represented in Figure \ref{Fig_SWT_Blocks}. Similar to downsampling in CNN, the effect of patch merging is to half the feature map's size while doubling the channels, which helps produce a hierarchical representation. The swin-transformer block is a module for feature extraction which is described in Formula \ref{Eq_SWT1}-\ref{Eq_SWT2}.

\begin{align} \label{Eq_SWT1}
\hat{X}^l = W{\textendash}SMA(LN(X^{l-1})) + X^{l-1}, \\
X^l = MLP(LN(\hat{X}^l)) + \hat{X}^l, \\
\hat{X}^{l+1} = SW{\textendash}SMA(LN(X^l)) + X^l, \\
\label{Eq_SWT2}
X^{l+1} = MLP(LN(\hat{X}^{l+1})) + \hat{X}^{l+1}
\end{align} 
where $X^l$ → $l$th-layer feature map, and 

\ \ \ \ \ $LN$ → layer normalization. 

\subsection{Feature extraction strategy}

\begin{figure}
  \centering
  \includegraphics[width=0.85\linewidth]{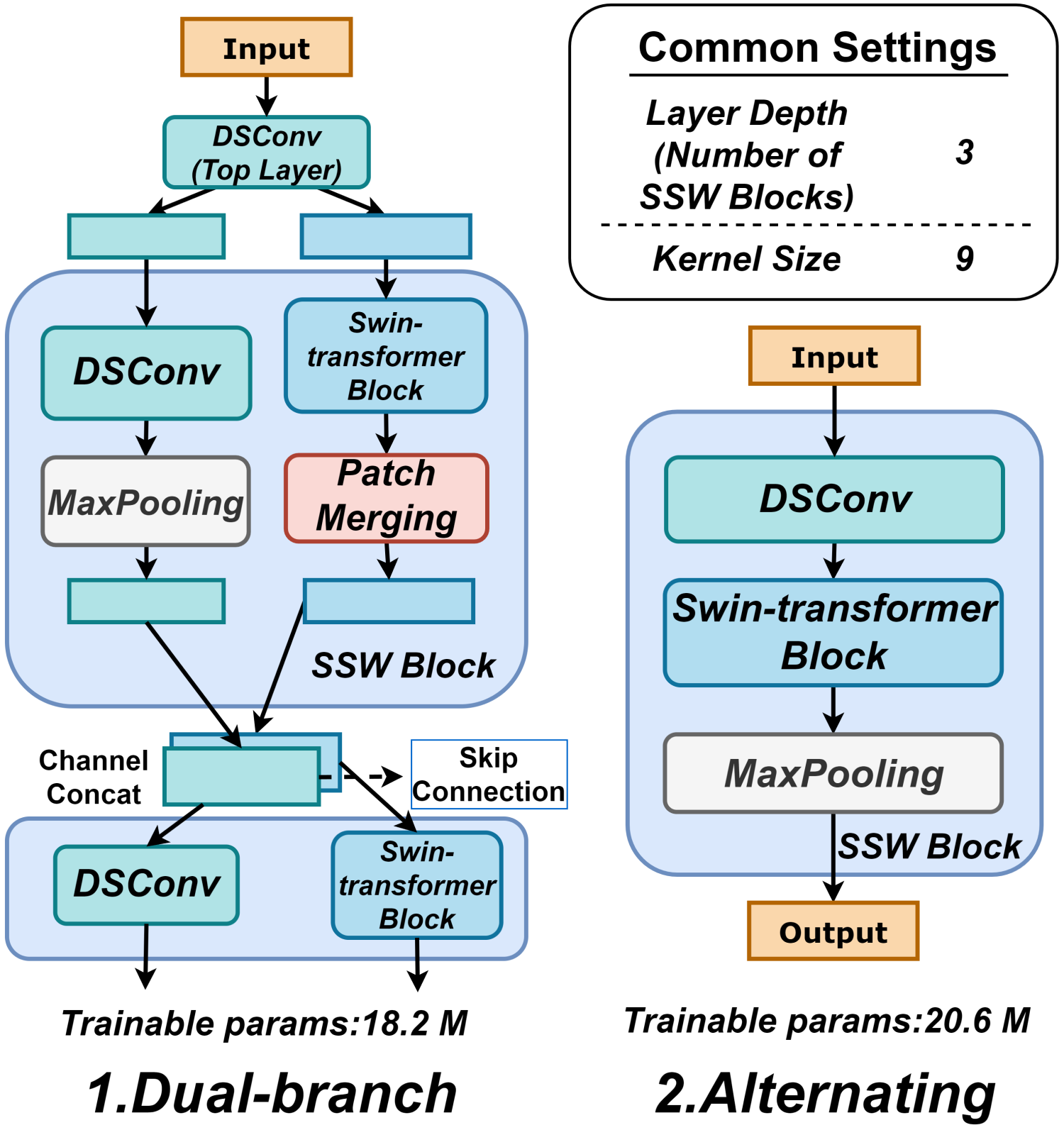}
  \caption{Two model architecture types are designed to explore better feature extraction strategy for SSW-OCTA.}
  \label{Fig_SSW_Arch}
\end{figure}

To explore an optimal strategy for extracting local and global vessel features, we set up two types of network architectures, namely dual-branch and alternating, which are illustrated in Figure \ref{Fig_SSW_Arch}. We first fixed the key hyperparameters of the networks, such as kernel size, number of channels, and layer depth, to ensure a fair comparison under comparable parameter counts. The model parameters count is primarily determined by the number of network layers and the initial layer's channel size. In the downsampling feature extraction process, the feature map channels double, and the spatial size halves as passing every SSW block. For the dual-branch and alternating architectures, we set the initial channel sizes to 72 and 108, respectively. After the better architecture has been selected, the key hyperparameters are subjected to ablation experiments to assess their effect on model performance. This provides a quantitative analysis of different segmentation tasks and contributes to the model's lightweight.

\section{Experiments} \label{Sec_Experiments}

\subsection{Dataset and experimental settings} \label{Subsec_Setting}

The publicly available dataset used for experiments in this paper is OCTA-500 ~\cite{li2020ipn}. This dataset is divided into two sub-datasets based on the field of view (Fov): 3M ($3mm \times 3mm$) and 6M ($6mm \times 6mm$). In the provided en-face images, the resolution for the 3M subset is $304 \times 304$ with 200 samples, and for the 6M subset, it is $400 \times 400$ with 300 samples. OCTA-500 has the largest sample size and the most comprehensive annotations, including RV, FAZ, capillary, artery, and vein. For ease of comparison, the providers of the OCTA-500 dataset proposed a data partitioning method for model evaluation, as shown in Table \ref{Table_Partition}. The input for training and testing consists of en-face OCTA images from the FULL, ILM (internal limiting membrane)\_OPL (outer plexiform layer), and OPL\_BM (Bruch’s membrane).

\begin{table}[t]
\centering
\caption{The OCTA-500 Dataset Partition (by Sample ID).}
\label{Table_Partition}
\begin{tabular}{cccc}
\toprule
FoV & Training Set & Validation Set & Test Set \\
\midrule
3M & 10301-10440 & 10441-10450 & 10451-10500 \\
6M & 10001-10180 & 10181-10200 & 10200-10300 \\
\bottomrule
\end{tabular}
\end{table}

The SSW-OCTA model is deployed on a 4090Ti GPU with 24GB of memory. The optimizer used is AdamW, and a warm-up strategy is employed for the learning rate during training, starting from $10^{-4}$ and gradually increasing to $10^{-2}$. The number of training epochs is set to 100 and the results are recorded every 10 epochs. The test set evaluation metrics are selected as outcomes when the validation set loss is minimized. The data augmentation strategy applied in the training set includes flipping, brightness adjustment, random slight rotation, blur transformations, and coarse dropout, implemented using the Albumentations toolkit \cite{info11020125}. The loss function is clDice Loss and can be represented as: 
\begin{align} \label{Eq_clDice}
L_{clDice} = 0.8 * L_{Dice} + 0.2 * L_{clDice}^\prime
\end{align} 

where $ L_{Dice} = 1 - \frac{2 * |\hat{Y} \cap Y|}{|\hat{Y}| + |Y|} $, 

\ \ \ \ \ $ L_{clDice}^\prime = 1 - 2 * \frac{T_{prec}(\hat{Y_s}, Y) * T_{sens}(Y_s, \hat{Y})}{T_{prec}(\hat{Y_s}, Y) + T_{sens}(Y_s, \hat{Y})} $, 

\ \ \ \ \ $Y$ → {\itshape the ground-truth}, 

\ \ \ \ \ $\hat{Y}$ → {\itshape the predicted value}, 

\ \ \ \ \ $Y_s, \hat{Y_s}$ → {\itshape soft{-}skeleton($Y$, $\hat{Y}$)}, and 

\ \ \ \ \ $T_{prec}, T_{sens}$ → {\itshape precision and sensitivity.}

The segmentation results using metrics Dice and Jaccard (JAC), which are calculated as follows:
\begin{align}
Dice(\hat{Y}, Y) = \frac{2 |\hat{Y} \cap Y|}{|\hat{Y}| + |Y|}
\end{align}

\begin{align}
Jaccard(\hat{Y}, Y) = \frac{|\hat{Y} \cap Y|}{|\hat{Y} \cup Y|}
\end{align}

\begin{table*}[t]
\centering
\caption{Ablation Results of Three Types of Model Architectures on OCTA-500 Tasks.}
\label{Table_Architectures}
\begin{tabular}{cccccccccccc}
\toprule
\multirow{2}{*}{FoV} & \multirow{2}{*}{Architecture} & \multicolumn{2}{c}{RV} & \multicolumn{2}{c}{FAZ} & \multicolumn{2}{c}{Capillary} & \multicolumn{2}{c}{Artery} & \multicolumn{2}{c}{Vein} \\
\cmidrule(lr){3-4}\cmidrule(lr){5-6}\cmidrule(lr){7-8}\cmidrule(lr){9-10}\cmidrule(lr){11-12}
\multicolumn{2}{c}{} & DICE ↑ & JAC ↑ & DICE ↑ & JAC ↑ & DICE ↑ & JAC ↑ & DICE ↑ & JAC ↑ & DICE ↑ & JAC ↑\\

\midrule
\multirow{3}{*}{3M}
& Dual-branch & \textbf{\underline{0.9215}} & \textbf{\underline{0.8553}} & \textbf{\underline{0.9816}} & \textbf{\underline{0.9641}} & \textbf{\underline{0.9051}} & \textbf{\underline{0.8271}} & \textbf{\underline{0.9047}} & \textbf{\underline{0.8269}} & \textbf{\underline{0.8903}} & \textbf{\underline{0.8035}} \\
& Alternating & 0.9212 & 0.8547 & 0.9764 & 0.9544 & 0.9041 & 0.8256 & 0.9017 & 0.8222 & 0.8831 & 0.7894 \\
& DSCNet & 0.9200 & 0.8526 & 0.9775 & 0.9565 & 0.9032 & 0.8239 & 0.8948 & 0.8108 & 0.8803 & 0.7878 \\
\midrule
\multirow{3}{*}{6M}
& Dual-branch & \textbf{\underline{0.8946}} & \textbf{\underline{0.8102}} & \textbf{\underline{0.9042}} & \textbf{\underline{0.8416}} & \textbf{\underline{0.8782}} & \textbf{\underline{0.7836}} & \textbf{\underline{0.8667}} & \textbf{\underline{0.7669}} & 0.8662 & 0.7675 \\
& Alternating & 0.8943 & 0.8097 & 0.8954 & 0.8293 & 0.8765 & 0.7812 & 0.8647 & 0.7640 & \textbf{\underline{0.8670}} & \textbf{\underline{0.7682}} \\
& DSCNet & 0.8914 & 0.8048 & 0.8992 & 0.8338 & 0.8725 & 0.7749 & 0.8540 & 0.7477 & 0.8536 & 0.7474 \\
\bottomrule
\end{tabular}
\end{table*}
\begin{table*}[t]
\centering
\caption{Segmentation Performance Metrics on Multiple OCTA-500 Tasks.}
\label{Table_Training}
\begin{tabular}{cccccccccccc}
\toprule
\multirow{2}{*}{Methods} & \multirow{2}{*}{FoV} & \multicolumn{2}{c}{RV} & \multicolumn{2}{c}{FAZ} & \multicolumn{2}{c}{Capillary} & \multicolumn{2}{c}{Artery} & \multicolumn{2}{c}{Vein} \\
\cmidrule(lr){3-4}\cmidrule(lr){5-6}\cmidrule(lr){7-8}\cmidrule(lr){9-10}\cmidrule(lr){11-12}
\multicolumn{2}{c}{} & DICE ↑ & JAC ↑ & DICE ↑ & JAC ↑ & DICE ↑ & JAC ↑ & DICE ↑ & JAC ↑ & DICE ↑ & JAC ↑\\

\midrule
\multirow{2}{*}{UNet}
& 3M & 0.9125 & 0.8398 & 0.9708 & 0.9440 & 0.8968 & 0.8136 & 0.8655 &  0.7647 & 0.8439 & 0.7316 \\
& 6M & 0.8854 & 0.7952 & 0.8718 & 0.7911 & 0.8608 & 0.7566 & 0.8382 & 0.7243 & 0.8419 & 0.7300 \\
\cmidrule(lr){2-12}
\multirow{2}{*}{SwinUNetR}
& 3M & 0.9180 & 0.8492 & 0.9780 & 0.9574 & 0.9037 & 0.8248 & 0.8887 & 0.8008 & 0.8796 & 0.7862 \\
& 6M & 0.8919 & 0.8058 & 0.9046 & 0.8399 & 0.8734 & 0.7763 & 0.8521 & 0.7448 & 0.8556 & 0.7503 \\
\cmidrule(lr){2-12}
\multirow{2}{*}{SegResNet}
& 3M & 0.9030 & 0.8238 & 0.9732 & 0.9484 & 0.8989 & 0.8167 & 0.8631 & 0.7603 & 0.8504 & 0.7405 \\
& 6M & 0.8855 & 0.7952 & 0.9095 & 0.8482 & 0.8639 & 0.7614 & 0.8296 & 0.7104 & 0.8287 & 0.7094 \\
\cmidrule(lr){2-12}
\multirow{2}{*}{FlexUNet}
& 3M & 0.9170 & 0.8474 & 0.9758 & 0.9533 & 0.9033 & 0.8239 & 0.8783 & 0.7842 & 0.8725 & 0.7746 \\
& 6M & 0.8888 & 0.8006 & 0.9094 & 0.8450 & 0.8683 & 0.7679 & 0.8573 & 0.7518 & 0.8544 & 0.7476 \\
\cmidrule(lr){2-12}
\multirow{2}{*}{DiNTS}
& 3M & 0.9165 & 0.8465 & 0.9769 & 0.9554 & 0.9035  & 0.8244 & 0.8868 & 0.7977 & 0.8721 & 0.7741 \\
& 6M & 0.8896 & 0.8018 & 0.9063 & 0.8479 & 0.8708 & 0.7719 & 0.8628 & 0.7606 & 0.8650 & 0.7643 \\
\midrule
\multirow{2}{*}{\textbf{SSW-OCTA}}
& 3M & 0.9215 & 0.8553 & 0.9816 & 0.9641 & 0.9051 & 0.8271 & 0.9047 & 0.8269 & 0.8903 & 0.8035 \\
& 6M & 0.8946 & 0.8102 & 0.9042 & 0.8416 & 0.8782 & 0.7836 & 0.8667 & 0.7669 & 0.8662 & 0.7675 \\
\bottomrule
\end{tabular}
\end{table*}
\begin{table}[t]
\centering
\caption{RV and FAZ Segmentation Results on OCTA-500 (underscores indicate the top two highest values).}
\setlength{\extrarowheight}{-2pt}
\label{Table_Comparison}
\begin{tabular}{cccccc}
\toprule
\multirow{2}{*}{Methods} & \multirow{2}{*}{Label} & \multicolumn{2}{c}{3M} & \multicolumn{2}{c}{6M} \\
\cmidrule(lr){3-4} \cmidrule(lr){5-6}
\multicolumn{2}{c}{} & DICE ↑ & JAC ↑ & DICE ↑ & JAC ↑ \\
\midrule
FARGO       & RV  & 0.9168 & 0.8470 & 0.8915 & 0.8050 \\
(2021)      & FAZ & \underline{0.9839} & \underline{0.9684} & \underline{0.9272} & \underline{0.8701} \\
\cmidrule(lr){3-4} \cmidrule(lr){5-6}
Joint-Seg   & RV  & 0.9113 & 0.8378 & \underline{0.8972} & \underline{0.8117} \\
(2022)      & FAZ & \underline{0.9843} & \underline{0.9693} & 0.9051 & 0.8424 \\
\cmidrule(lr){3-4} \cmidrule(lr){5-6}
RPS-Net     & RV  & 0.9155 & 0.8448 & 0.8989 & 0.8173 \\
(2022)      & FAZ & 0.9780 & 0.9582 & 0.9140 & \underline{0.8530} \\
\cmidrule(lr){3-4} \cmidrule(lr){5-6}
OCT2Former  & RV  & 0.9193 & 0.8513 & 0.8945 & 0.8099 \\
(2023)      & FAZ & -      & -      & -      & -      \\
\cmidrule(lr){3-4} \cmidrule(lr){5-6}
ASDC        & RV  & -      & -      & 0.8664 & 0.7904 \\
(2023)      & FAZ & -      & -      &\underline{0.9160} & 0.8520 \\
\cmidrule(lr){3-4} \cmidrule(lr){5-6}
DB-UNet     & RV  & 0.9180 & 0.9890 & 0.8931 & 0.8074 \\
(2023)      & FAZ & -      & -      & -      & -      \\
\cmidrule(lr){3-4} \cmidrule(lr){5-6}
SAM-OCTA    & RV  & \underline{0.9199} & \underline{0.8520} & 0.8869 & 0.7975 \\
(2024)      & FAZ & 0.9838 & 0.9692 & 0.9073 & 0.8473 \\
\cmidrule(lr){1-6}
SSW-OCTA    & RV  & \underline{\textbf{0.9215}} & \underline{\textbf{0.8553}} & \underline{\textbf{0.8946}} & \underline{\textbf{0.8102}} \\
(ours)      & FAZ & 0.9816 & 0.9641 & 0.9155 & 0.8608 \\
\bottomrule
\end{tabular}
\end{table}

\subsection{Ablations} \label{Subsec_Ablations}

We first tested the two designed model architectures on annotated OCTA-500 enface segmentation tasks and compared them with a baseline DSCNet. The results are presented in Table \ref{Table_Architectures}. Both SSW-OCTA architectures outperform the baseline DSCNet in terms of vessel segmentation performance. The dual-branch architecture generally surpasses the alternating one in most tasks. The alternating architecture interferes with the advantage of the DSCNet's local feature extraction, resulting in poor performance in FAZ segmentation. Therefore, the dual-branch strategy is chosen as the new baseline for the key hyperparameter ablation experiments.

\begin{figure}
  \centering
  \includegraphics[width=1\linewidth]{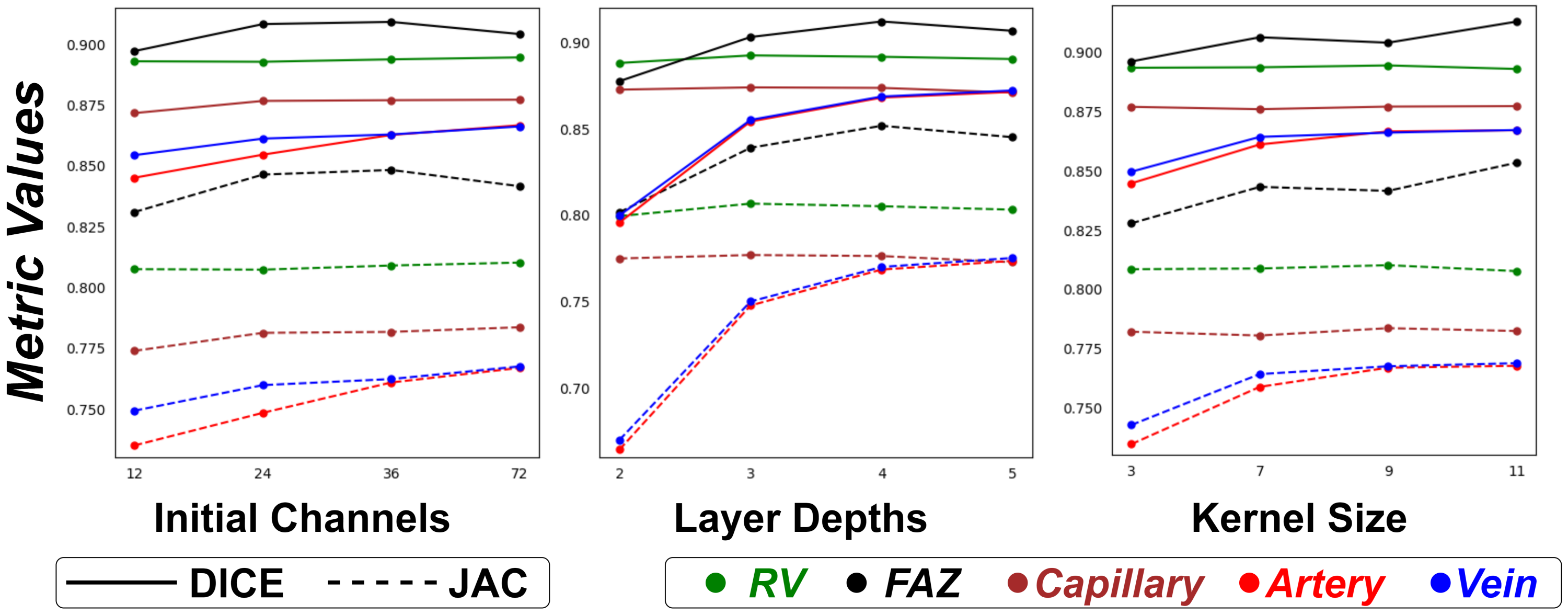}
  \caption{Ablation results of the SSW-OCTA in dual-branch architecture on the OCTA-500 6M subset.}
  \label{Fig_Hyper}
\end{figure}

Further ablation experiments are conducted with the channel number, network layer, and kernel size. For clarity, only the result of the 6M subset is summarized in Figure \ref{Fig_Hyper}, which is able to elucidate the effect of these hyperparameters. Reducing the initial channel size has a minimal impact on the RV and capillary segmentation and slightly decreases the performance in the FAZ, artery, and vein segmentation. Beyond an initial channel size of 36, the model's performance plateaus or even slightly declines, showing negligible further improvement. For the layer depth, the initial channel size is set to a small value of 12 to magnify its impact on model performance. With shallower layer depths, RV and capillary segmentation are unaffected, while performance for the FAZ dips slightly, and AV segmentation declines more significantly. The kernel size can similarly affect model performance, and larger kernel sizes notably improve FAZ segmentation performance. As the kernel size increases, the computational complexity also almost linearly increases. This ablation result suggests that a lightweight model suffices for RV and capillary segmentation, while for FAZ and AV segmentation, increasing the channels, layers, and kernel size is necessary for optimal results.

\subsection{Comparison results}

To investigate the performance of our model across diverse tasks, we trained and tested the following models: UNet, SwinUNetR, SegResNet, FlexUNet, DiNTS, and our SSW-OCTA using the dual-branch configures in Section \ref{Subsec_Ablations} \cite{ronneberger2015u, hatamizadeh2021swin, myronenko20193d, he2021dints}. As general segmentation models, the versions of the adopted models are implemented by the MONAI library \cite{cardoso2022monai}. For other OCTA-specific models, we selected the following representative 2D en-face input models: FARGO\cite{peng2021fargo}, Joint-Seg\cite{hu2022joint}, RPS-Net\cite{li2022rps}, OCT2Former\cite{tan2023oct2former}, ASDC\cite{khan2023adaptive}, DB-UNet\cite{wang2023db}, and SAM-OCTA\cite{wang2023sam}. The experimental designs of these works are comprehensive and have demonstrated promising results in RV and FAZ segmentation tasks. To showcase the competitiveness of our proposed model, we utilized the optimal hyperparameters and set the kernel size to 15 for the FAZ of the 6M subset. The experimental value results are presented in Table \ref{Table_Training} and \ref{Table_Comparison}.

In contrast to general segmentation models, SSW-OCTA excels in all segmentation tasks for the OCTA-500 dataset. The segmentation results for the capillary and AV are notable because previous research rarely comprehensively summarized them, which could provide references for consequence studies. Attention mechanisms play a vital role in vessel segmentation, particularly for AV. This is evidenced by the impressive metrics achieved by SwinUNetR and SSW-OCTA. Targets in vessel tasks are distributed across the entire image with varying quantities, and the attention mechanism proves effective in capturing interrelationships across non-adjacent regions. Although the improvement is limited for RV and capillary tasks, it benefits more for the segmentation of arteries and veins. This arises because the arteries and veins, being mutually exclusive subsets of the RV in the OCTA-500 dataset annotations, exhibit similarities in shape and distribution patterns. The model not only needs to capture vessel shape features but also discern vessel types, highlighting the necessity for the model's ability to integrate global information. In OCTA segmentation tasks of vessel types, the introduction of suitable attention mechanisms could avoid performance degradation or even have improvements. 

\begin{figure}
  \centering
  \includegraphics[width=1\linewidth]{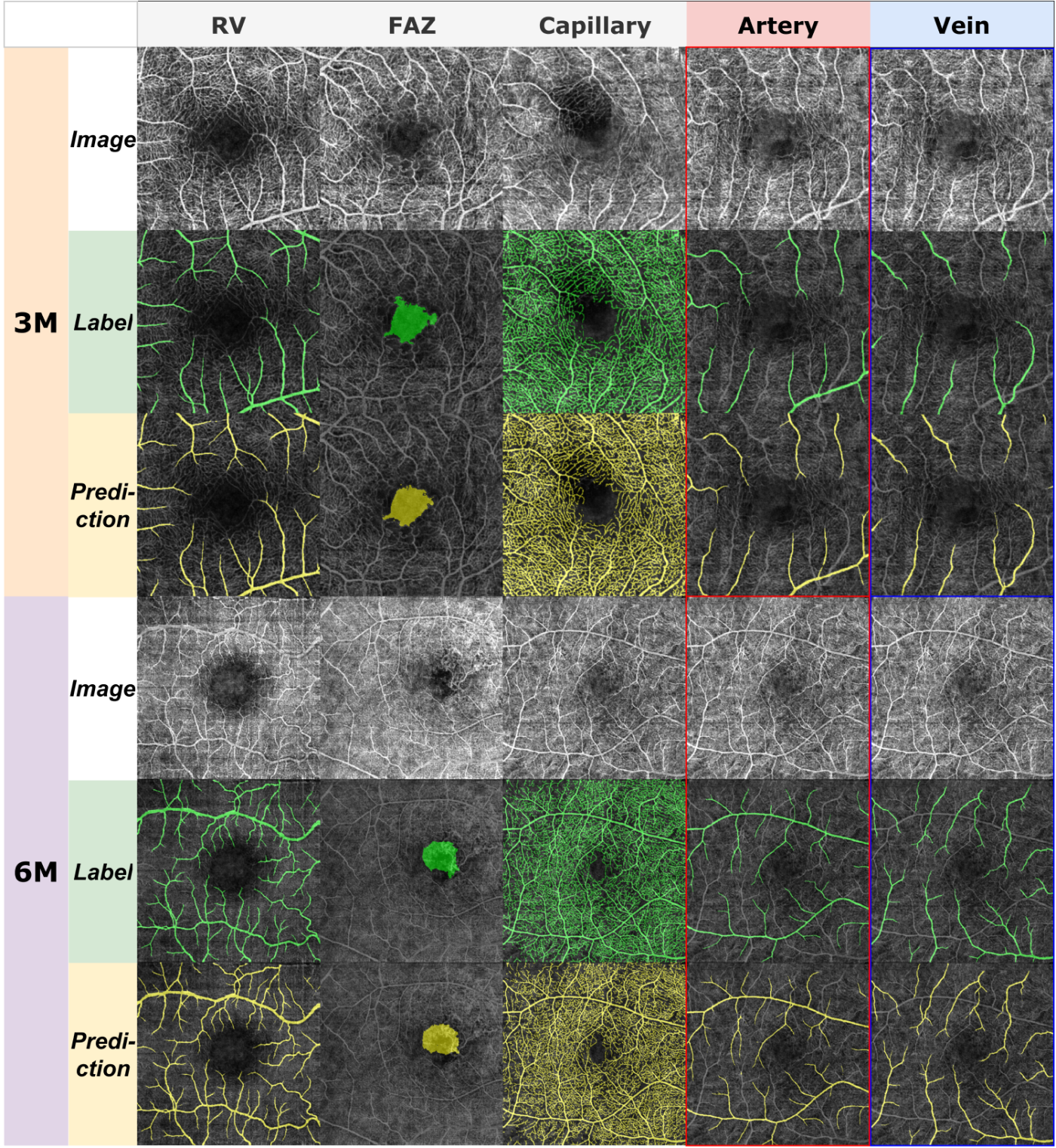}
  \caption{Examples of SSW-OCTA segmented on OCTA-500 tasks. The artery and vein tasks are selected from the same sample, and their borders are colored in red and blue, respectively, for clear distinction and comparison.}
  \label{Fig_Result}
\end{figure}

Compared to other SOTA models, our SSW-OCTA model achieves comparable scores in metrics and even stands out in RV segmentation. Due to approaching a performance bottleneck on the specific dataset, the scores of these models are very close, while more intuitive results can be observed in Figure \ref{Fig_Result}. In tasks of RV, capillary, and AV segmentation, the SSW-OCTA demonstrates accuracy in maintaining continuity and morphology in large vessel targets. The segmentation of the vessel stem shape is highly precise, with only minimal errors observed in small branches. For capillary, it effectively reflects the density of different regions in the image, and it can reasonably distinguish between arteries and veins in AV segmentation. These results demonstrate the effectiveness and advancement of our model. A limitation lies in the FAZ shape segmentation error, particularly in the outward protruding parts of the edge, more in 3M samples. The inference is that the FAZ occupies a relatively fixed region in the image's center, and its quantity is unique, making the target-associated image context more localized. This facilitates the convolutional block's inductive bias feature. 

\section{Conclusion} \label{Sec_Conclusion}

This paper proposes the SSW-OCTA segmentation model based on the shape and distribution characteristics of vessels in OCTA retinal images. It is capable of adapting to tubular morphology and perceiving global distribution. The model design incorporates the step-offset deformable convolution kernel technique from the DSCNet and the windowed attention mechanism from the swin-transformer. The model achieves high-performance segmentation of RV and capillary in en-face OCTA images with a lightweight version (approximately 170k trainable parameters). It also shows promising performance in FAZ and AV segmentation tasks.



\section*{ACKNOWLEDGMENT}

This work is supported by the Chongqing Natural Science Foundation Innovation and Development Joint Fund (CSTB2023NSCQ-LZX0109), Chongqing Technology Innovation \& Application Development Key Project (cstb2022tiad-kpx0148), and Fundamental Research Funds for the Central Universities (No.2022CDJYGRH-001).

\bibliographystyle{IEEEtrans}
\bibliography{smc2024}

\end{document}